\documentclass{ieeetj}
\usepackage{cite}
\usepackage{algorithmic}
\usepackage{graphicx,color}
\usepackage{textcomp}
\usepackage{xcolor}
\usepackage{hyperref}
\usepackage{booktabs}
\usepackage{multirow}
\usepackage[table]{xcolor}
\hypersetup{hidelinks=true}
\def\BibTeX{{\rm B\kern-.05em{\sc i\kern-.025em b}\kern-.08em
    T\kern-.1667em\lower.7ex\hbox{E}\kern-.125emX}}
\AtBeginDocument{\definecolor{tmlcncolor}{cmyk}{0.93,0.59,0.15,0.02}\definecolor{NavyBlue}{RGB}{0,86,125}}

\usepackage{amsmath,amsfonts}
\usepackage{array}
\usepackage[caption=false,font=normalsize,labelfont=sf,textfont=sf]{subfig}
\usepackage{textcomp}
\usepackage{stfloats}
\usepackage{verbatim}
\usepackage{graphicx}
\usepackage{cite}
\usepackage{tabularx}
\usepackage{booktabs}

\usepackage{acronym}
\acrodef{MIMO}{multiple-input multiple-output}
\acrodef{RF}{radio frequency}
\acrodef{LoS}{line-of-sight}
\acrodef{NLoS}{non-line-of-sight}
\acrodef{AoA}{angle-of-arrival}
\acrodef{AoD}{angle-of-departure}
\acrodef{UPA}{uniform planar array}
\acrodef{ARV}{array response vector}
\acrodef{EM}{electromagnetic}
\acrodef{ER-FAS}{electromagnetically reconfigurable fluid antenna system}
\acrodef{SV}{Saleh-Valenzuela}

\usepackage{color} 
\newcommand{\red}[1]{{\color{red}{#1}}}

\usepackage{tikz,pgfplots}
\usepackage{amsmath,amssymb}
\pgfplotsset{compat=newest}
\usepackage{tabu,longtable}
\usepackage{diagbox}
\usepackage{threeparttable}
\usepackage{makecell}
\usepackage{multirow} 
\usepackage{units} 		
\usepackage{bm} 		
\usepackage{amssymb} 	

\newcommand{\HH}{\mathsf{H}}

\newcommand{\av}{{\bf a}}

\newcommand{\wv}{{\bf w}}

\newcommand{\xv}{{\bf x}}
\newcommand{\yv}{{\bf y}}

\newcommand{\Hm}{{\bf H}}

\newcommand{\Sm}{{\bf S}}

\newcommand{\It}{{\rm I}}

\newcommand{\Rt}{{\rm R}}

\newcommand{\Tt}{{\rm T}}

\newcommand{\phiv}{\hbox{\boldmath$\phi$}}

\newcommand{\Gammam}{\hbox{\boldmath$\Gamma$}}

\usepackage{hyperref} 	

\def\OJlogo{\vspace{-4pt}}
\def\seclogo{\vspace{10pt}$$$$}

\def\authorrefmark#1{\ensuremath{^{\textbf{#1}}}}

\begin{document}

\receiveddate{This work has been submitted to the IEEE for possible publication. Copyright may be transferred without notice, after which this version may no longer be accessible.}


\title{Millimeter-Wave RIS: Hardware Design and System-Level Considerations}

\author{Ruiqi Wang\authorrefmark{1}, Pinjun Zheng\authorrefmark{2}, Member, IEEE, Yiming Yang\authorrefmark{1}, Xiarui Su\authorrefmark{1}, Mohammad Vaseem\authorrefmark{1}, Anas Chaaban\authorrefmark{2}, Senior Member, IEEE, Md. Jahangir Hossain\authorrefmark{2}, Senior Member, IEEE, Tareq Y. Al-Naffouri\authorrefmark{1}, Fellow, IEEE, and Atif Shamim\authorrefmark{1}, Fellow, IEEE}

\affil{Computer, Electrical and Mathematical Sciences and Engineering (CEMSE), King Abdullah University of Science and Technology~(KAUST), Thuwal 23955-6900, Kingdom of Saudi Arabia}
\affil{School of Engineering, The University of British Columbia, Kelowna, BC V1V 1V7, Canada }
\corresp{Corresponding author: Ruiqi Wang (email: ruiqi.wang.1@kaust.edu.sa).}

\begin{abstract}
Reconfigurable intelligent surfaces have emerged as a promising hardware platform for shaping wireless propagation environments at millimeter-wave (mm-Wave) frequencies and beyond. While many existing studies emphasize channel modeling and signal processing, practical RIS deployment is fundamentally governed by hardware design choices and their system-level implications. This paper presents a hardware-centric overview of recent mm-Wave RIS developments, covering wideband realizations, high-resolution phase-quantized designs, fully printed low-cost implementations, optically transparent surfaces, RIS-on-chip solutions, and emerging three-dimensional architectures. Key challenges including mutual coupling, calibration, multi-RIS interaction, and frequency-dependent phase control are discussed to bridge hardware realization with system-level optimization. This overview provides practical design insights and aims to guide future RIS research toward scalable, efficient, and practically deployable intelligent surface architectures.
\end{abstract}


\IEEEspecialpapernotice{(Invited Article)}

\maketitle

\section{INTRODUCTION}
\IEEEPARstart{M}{illimeter-wave} (mm-Wave) communication has been widely recognized as a key enabler for fifth-generation (5G) and future sixth-generation (6G) wireless systems, owing to its abundant spectral resources and potential to support extremely high data rates~\cite{Jay20256G},~\cite{Hongwei2021mmWave},~\cite{Xue2024mmWaveTHz},~\cite{Moltchanov2022mmWave},~\cite{Chukhno2024mmWave},~\cite{Ruiqi2025MIMO}. By exploiting carrier frequencies typically ranging from tens of gigahertz (GHz) to sub-terahertz (sub-THz), mm-Wave systems can provide multi-gigabit throughput, which is essential for emerging applications such as immersive extended reality (XR)~\cite{Sampath2024XR}, high-resolution sensing, and wireless backhaul and fronthaul links~\cite{Rappaport20196G}. However, the practical deployment of mm-Wave communication systems faces fundamental challenges stemming from severe path loss, high penetration attenuation, and strong susceptibility to blockage by common obstacles~\cite{Ju2021mmWave}. These limitations significantly restrict coverage and reliability, particularly in dense urban and indoor environments, and have led to ongoing discussions regarding the large-scale practicality and cost-effectiveness of mm-Wave deployments.

Reconfigurable Intelligent Surfaces (RIS) have recently attracted significant research interest as a potential means to mitigate some of these propagation limitations by enabling programmable manipulation of the wireless environment~\cite{DiRenzo2020RIS}. An RIS is typically composed of a large number of subwavelength passive or active elements whose electromagnetic (EM) responses can be dynamically adjusted to control the phase, amplitude, or polarization of incident waves~\cite{Cheng2022Proceeding}. By intelligently configuring these elements, RIS can reshape propagation paths, enhance signal coverage in blockage-prone or non-line-of-sight scenarios, suppress interference, and improve spectral and energy efficiency~\cite{Huang2019RIS}, ~\cite{Tang2022RIS}. Owing to their passive operation and potential for compact implementation, as well as their compatibility with existing wireless infrastructure, RIS have attracted substantial attention as a key building block for 5G-Advanced and 6G systems~\cite{Rasilainen2023RIS}.

Although RIS technology has been extensively investigated in academia, questions remain regarding its practical deployment maturity and industry acceptance. Recently, initial real-world pilot deployments have started to emerge. For example, China Mobile Hubei Company publicly announced the procurement of approximately 300 reconfigurable intelligent surfaces for 5G network enhancement, targeting coverage improvement in challenging non-line-of-sight scenarios and dense urban environments~\cite{cmcc_hubei_ris_300_2025}. This procurement reflects exploratory industrial interest in RIS-assisted coverage optimization. In addition, operator--vendor verification activities in 5G-Advanced contexts have demonstrated the feasibility of integrating RIS into real network environments~\cite{zte_cmcc_ris_2023}. Nevertheless, RIS is currently deployed primarily in trial or limited commercial scenarios rather than in fully standardized large-scale rollouts. Several factors contribute to this cautious adoption, including installation complexity, calibration requirements, long-term reliability concerns, cost-performance trade-offs, and the absence of mature standardization frameworks. From a standardization perspective, RIS is being discussed in 3GPP studies mainly in the context of enhancing secondary links or improving non-line-of-sight coverage. Comprehensive integration into mainstream wireless standards has not yet been realized and remains under active investigation.

\subsection{Motivation for mm-Wave RIS in 5G and 6G Communications}

The motivation for integrating RIS into mm-Wave communication systems is multifaceted. One primary motivation lies in the inherent propagation characteristics of mm-Wave signals, including limited diffraction capability, high penetration loss, and strong sensitivity to blockage by common obstacles such as buildings, foliage, and the human body. These characteristics often result in fragmented coverage, frequent link outages, and severe performance degradation in non-line-of-sight scenarios. Conventional approaches, such as dense base-station deployment, active relays can partially alleviate these issues. However, these solutions rely on active radio-frequency chains and additional infrastructure, and their relative cost-performance tradeoffs compared to RIS-assisted solutions remain scenario-dependent and subject to ongoing investigation~\cite{DiRenzo2020RISVSRelay}. In contrast, RIS offers a complementary and potentially cost-efficient solution in specific deployment scenarios by reshaping the wireless propagation environment. By intelligently redirecting incident signals toward desired directions, RIS can enhance coverage and improve link reliability without introducing active amplification or significant energy overhead, making them particularly attractive for targeted coverage enhancement and blockage mitigation.

Beyond coverage enhancement, another important motivation for mm-Wave RIS integration is the evolving role of future 6G networks, which are expected to support not only high-rate communications but also integrated sensing, localization, and environmental awareness~\cite{Tataria6G}. The capability of RIS to dynamically control EM wavefronts enables flexible manipulation of signal reflection, transmission, and scattering, which is highly beneficial for multifunctional operation. At mm-Wave frequencies, the short wavelength allows for fine-grained spatial resolution and precise wavefront shaping using electrically large RIS apertures, thereby facilitating high-angular-resolution sensing and accurate positioning. As a result, RIS can act as a shared physical-layer platform that jointly supports communication and sensing functionalities, contributing to the realization of smart radio environments in which the propagation medium itself becomes programmable, adaptive, and context-aware. This paradigm shift is widely regarded as a key enabler for future 6G systems, where tight integration of communication, sensing, and intelligence is expected to play a central role.

\subsection{RIS Hardware Design and System-level Challenges at mm-Wave Frequencies}

Despite their promising potential, the practical realization of RIS at mm-Wave frequencies poses significant design and implementation challenges. From a hardware perspective, material selection and fabrication accuracy become critical due to the small feature sizes required at high frequencies. Conductor and dielectric losses are more pronounced in the mm-Wave regime, which can severely degrade RIS efficiency if not properly addressed. Achieving wideband operation is another major challenge, as many RIS designs inherently exhibit narrowband responses caused by resonant element behavior.

Scalability and control complexity further complicate mm-Wave RIS implementations. Large-scale RIS arrays require efficient control architectures capable of configuring hundreds or thousands of elements with low latency and minimal overhead. Mutual coupling between closely spaced elements becomes non-negligible, affecting both EM performance and system-level modeling accuracy. In addition, practical deployment considerations, such as calibration, robustness to fabrication tolerances, and integration with existing communication and sensing platforms, remain open research problems.


Beyond hardware realization and industrial deployment considerations, the practical deployment of mm-Wave RIS is shaped by several system-level constraints that directly affect channel modeling and performance evaluation. Wideband operation, which is intrinsic to mm-Wave and future 6G systems, challenges the conventional narrowband assumption widely adopted in RIS analysis. In realistic OFDM-based links, propagation delays and the inherently frequency-dependent phase response of resonant unit cells lead to dispersive cascaded channels, rendering frequency-flat and perfectly separable RIS models only approximate. As a result, accurate system design should account for phase--frequency coupling and wideband-aware optimization.

RIS behavior is also strongly coupled with network-level interactions. In multi-RIS or dense deployments, the additional controllable propagation paths created by distributed surfaces may enhance desired links while simultaneously reshaping interference patterns. The coordination among multiple RIS panels, access points, and users becomes increasingly complex, particularly when surfaces are partitioned to support multi-functional operation or when reflective and transmissive modes coexist. These interactions require interference-aware configuration strategies and scalable control mechanisms to prevent interference-limited performance.

Moreover, hardware nonidealities propagate to the system level. Mutual coupling alters the effective reflection matrix and challenges the commonly assumed diagonal RIS response, while phase quantization and amplitude--phase coupling affect beamforming fidelity. Calibration errors and geometry uncertainty further introduce model mismatch, which can impact channel estimation, localization accuracy, and sensing reliability. These issues become more critical in emerging applications such as radiative near-field communication, integrated sensing and communication, and non-terrestrial networks, where large apertures, mobility, and stringent synchronization requirements tighten the coupling between EM design and system-level objectives. These challenges will be further examined in detail in subsequent sections, where both hardware and system-level implications are discussed.

\subsection{Objective and Organization}

In light of these challenges, a comprehensive understanding of mm-Wave RIS from both hardware and system perspectives is essential. This paper provides a comprehensive and hardware-centric overview of mm-Wave RIS, with a particular emphasis on practical design considerations and system-level implications. Unlike existing surveys that primarily focus on channel modeling, optimization algorithms, or abstract performance analysis, this work places mm-Wave RIS hardware realization at the core of the discussion and 
discusses the practical implications and open challenges at the interface between EM hardware design and communication system performance. The objective is to bridge the gap between theoretical mm-Wave RIS models and practical implementations, while examining the practical constraints in real-world RIS deployment, offering insights that contribute to the development of RIS-enabled 6G wireless systems.

The paper is organized to progressively guide the reader from fundamental concepts to advanced architectures and open challenges for mm-Wave RIS. Section II reviews the role of RIS in mm-Wave communications and sensing. Section III focuses on mm-Wave RIS hardware and implementation aspects, including wideband designs, phase quantization, printed and transparent RIS, on-chip implementations, and emerging 3D architectures. Section IV discusses key challenges in practical deployments. Finally, Section V outlines promising research directions toward scalable and multifunctional RIS technologies.

\section{The Role of RIS in mm-Wave Communication, Localization, and Sensing}

This section reviews the fundamental role of RIS in mm-Wave communication, localization, and sensing systems. 

\subsection{RIS-Aided Signal Transmission}

The functionality of the RIS lies in its capability to reflect incident EM waves with reconfigurable responses. In what follows, we present a commonly used signal model, which reveals the enabling role of RISs in mm-Wave communication, localization, and sensing. 

\begin{figure}[t]
  \centering
  \includegraphics[width=0.9\linewidth]{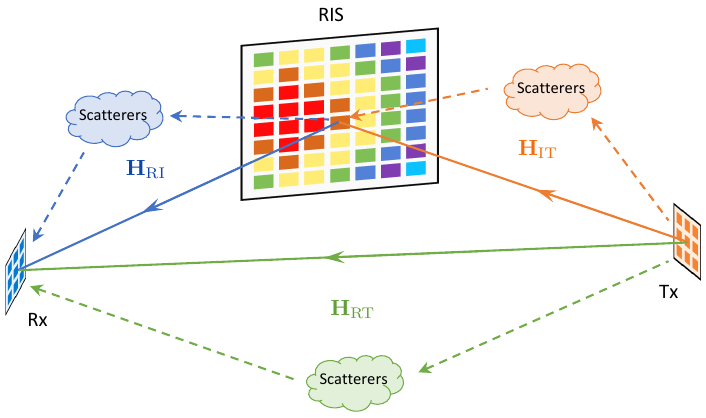}
  \caption{
    Schematic of the RIS-aided MIMO communication system. 
  }
\label{fig_system}
\end{figure}

Considering an mmWave MIMO system where an \(N_\Tt\)-antenna transmitter (Tx) sends signals to an \(N_\Rt\)-antenna receiver (Rx) through both a direct link and an RIS-assisted reflection link, as illustrated in Fig.~\ref{fig_system}, the signal transmission in the frequency-domain can be described as~\cite{Yang2024Near,Umer2025Reconfigurable}
\begin{equation}\label{eq:yfreq}
{\yv}_m[k]
= \big(\Hm_{\mathrm{RT}}[k]+\Hm_{\mathrm{RI}}[k]\Gammam_m\Hm_{\mathrm{IT}}[k]\big){\xv}_m[k]
+ {\wv}_m[k].
\end{equation}
where $k$ indexes the subcarrier, and $m$ denotes the symbol index.\footnote{The subscripts follow the matrix multiplication order of signal propagation, where ``$\mathrm{T}$'', ``$\mathrm{I}$'', and ``$\mathrm{R}$'' denote the Tx, RIS, and Rx, respectively.}
The vectors ${\xv}_m[k]\in\mathbb{C}^{N_\Tt}$ and ${\yv}_m[k]\in\mathbb{C}^{N_\Rt}$ represent the transmitted and received signal vectors at subcarrier $k$, respectively, while ${\wv}_m[k]$ denotes the additive noise at the receiver. The matrices $\Hm_{\mathrm{RT}}[k]\in\mathbb{C}^{N_\Rt\times N_\Tt}$, 
$\Hm_{\mathrm{IT}}[k]\in\mathbb{C}^{N_\It\times N_\Tt}$, and 
$\Hm_{\mathrm{RI}}[k]\in\mathbb{C}^{N_\Rt\times N_\It}$ denote the frequency-domain channel matrices of the direct Tx--Rx link, the Tx--RIS link, and the RIS--Rx link at subcarrier $k$, respectively, where $N_\It$ is the number of RIS reflecting elements. 
The diagonal matrix $\Gammam_m=\mathrm{diag}(\boldsymbol{\gamma}_m)\in\mathbb{C}^{N_\It\times N_\It}$ represents the RIS reflection coefficient matrix at symbol $m$. Specifically,  
$\boldsymbol{\gamma}_m=[\gamma_{m,1},\ldots,\gamma_{m,N_\It}]^\mathsf{T}$ and $\gamma_{m,n}$ is the response of the $n^{\text{th}}$ RIS unit cell, which can be expressed as
\begin{equation}\label{eq:Gammmam}
\gamma_{m,n}=a_{m,n} e^{\text{j}\omega_{m,n}},
\end{equation}
where $a_{m,n}$ and $\omega_{m,n}$ denotes the tunable amplitude and phase, respectively..

Accordingly, the effective cascaded channel at subcarrier $k$ is given by
\begin{equation}\label{eq:Heff}
    \Hm_{\mathrm{eff},m}[k]
= \Hm_{\mathrm{RT}}[k]+\Hm_{\mathrm{RI}}[k]\Gammam_m\Hm_{\mathrm{IT}}[k].
\end{equation}
The RIS response \(\Gammam_m\) may also be frequency-dependent in the wideband case~\cite{AbbasTAP2025,AbbasTWC2025}, although it is commonly assumed to be frequency-flat in the literature.


\subsection{Applications of RIS in mm-Wave Wireless Systems}

By properly configuring the reflection response, RISs can reshape the propagation environment and fundamentally alter the roles of multipath components in communication, localization, and sensing.

\subsubsection{Communication}

From~\eqref{eq:Heff}, the RIS can enhance the effective channel by introducing the cascaded channel $\Hm_{\mathrm{RI}}[k]\Gammam_m\Hm_{\mathrm{IT}}[k]$. This reflected link provides a reconfigurable copy of the transmitted signal, which can be exploited to enhance communication reliability and efficiency, or to enable communication when the direct link $\Hm_{\mathrm{RT}}[k]$ is severely blocked or unavailable.

Since the received signal explicitly depends on the RIS reflection matrix \(\Gammam\), the design of \(\Gammam\) has attracted extensive research interest. Existing studies consider a wide range of objectives, including maximizing spectral efficiency or achievable rate, minimizing bit error rate (BER), improving energy efficiency, extending coverage and link reliability, enhancing physical-layer security, reducing transmit power, and supporting dual-functional radar-communication (DFRC) performance~\cite{Huang2019Reconfigurable,ElMossallamy2020Reconfigurable,Ying2020GMD,Liu2022Joint,Zhang2023Active}. Beyond performance enhancement, RISs enable spatio-temporal-frequency control of mmWave propagation, shifting part of beamforming and coverage management from transceivers to the wireless channel~\cite{Renzo2020Smart}, which in turn has stimulated a broad range of analytical studies~\cite{Nadeem2020Asymptotic,Lu2022Outage,Liu2021Asymptotic}.

\subsubsection{Localization}

Compared to communication, RISs can play an even more transformative role in localization (or positioning). In addition to acting as a passive reflector, a RIS can serve as a synchronized reference point in the environment, generating additional propagation paths with known spatial signatures. To show this effect, one can express the RIS--Rx channel $\Hm_{\mathrm{RI}}[k]$ using a sparse geometric model as~\cite{Venugopal2017Channel,Heath2016Overview}
\begin{equation}
    \Hm_{\mathrm{RI}}[k] =  \sum_{i=1}^{L} \alpha_{i} e^{-\text{j}{2\pi (k-1)\Delta_f\tau_{i}}} \av_\Rt(f_k, \phiv_i)\av_\It^\HH(f_k, \bm{\theta}_i),
\end{equation}
where $L$ denotes the number of resolvable propagation paths between the RIS and the receiver, and $\alpha_i$ and $\tau_i$ represent the complex path gain and propagation delay of the $i^\text{th}$ path, respectively. 
The quantity $\Delta_f$ is the subcarrier spacing and $f_k$ is the passband frequency of the $k$th subcarrier. 
The vector $\av_\Rt(f_k,\phiv_i)\in\mathbb{C}^{N_\Rt}$ denotes the array response of the receiver toward the angle-of-arrival (AoA) $\phiv_i$ at frequency $f_k$, while $\av_\It(f_k,\bm{\theta}_i)\in\mathbb{C}^{N_\It}$ denotes the array response of the RIS toward the angle-of-departure (AoD) $\bm{\theta}_i$ at the same frequency. The expression of these array response vectors in the 3D space can be found in, e.g.,~\cite[Eq.~(2)]{Zheng2026Mutual}.

This parametric representation highlights that each RIS--Rx propagation path is characterized by a small set of physically meaningful parameters, including delay, angle, and complex gain. 
Since the RIS location and geometry are typically known and controllable, the associated angular signatures become predictable, effectively creating virtual anchors. As a result, the RIS introduces additional geometric constraints (e.g., AoA~$\phiv_i$) that can help identify the user position and orientation~\cite{Wymeersch2020Radio,He2022Beyond,Liu2021Reconfigurable}.

One notable advantage of RIS-assisted localization is its ability to enable localization in otherwise ill-posed or extreme scenarios. For instance, in a conventional single-input single-output (SISO) system consisting of only a base station (BS) and a user equipment (UE), communication may be feasible, but localization is generally impossible due to the lack of geometric diversity. By introducing a RIS and exploiting the delay and angular information of the reflected path, joint UE localization and synchronization become achievable~\cite{Keykhosravi2022RIS}. More strikingly, with appropriately deployed RISs, it is even possible to perform localization without relying on any BSs, which is referred to as sidelink positioning~\cite{Hui2024Multi,Ammous20253D}. Recent studies have demonstrated the versatility of RIS-assisted localization across a wide range of scenarios, including localization under user mobility~\cite{Keykhosravi2022RIS}, seamless indoor–outdoor positioning~\cite{Zheng2023LEO}, near-field localization in large-aperture mm-wave systems~\cite{Pan2023RIS}, and satellite-based user positioning~\cite{Wang2024Beamforming}. 

\subsubsection{Integrated Sensing and Communication}

Building upon the communication and localization capabilities enabled by RISs, recent studies have explored their role in integrated sensing and communication (ISAC), where the same reflected propagation paths are exploited for both information transmission and environmental sensing~\cite{Chepuri2023}. By appropriately configuring the RIS reflection coefficients, the reflected channel can be shaped to support reliable data delivery while maintaining sufficient geometric diversity for sensing, enabling a unified treatment of communication and sensing functionalities.

From a system design perspective, RISs introduce additional controllable degrees of freedom that fundamentally reshape the sensing--communication tradeoff in ISAC systems. RISs can generate virtual line-of-sight paths and enhance target illumination without requiring additional active radio-frequency chains, thereby improving both sensing reliability and communication robustness in non-line-of-sight conditions~\cite{Liu2023Integrated}. These capabilities allow transmit beamforming, RIS configuration, and receiver processing to be jointly optimized to satisfy communication quality-of-service constraints while enhancing sensing metrics such as detection probability, Cramér-Rao bound (CRB), and angular resolution~\cite{Liu2026RIS,Liu2024SNR}.

\section{mm-Wave RIS Hardware and Implementation}

\subsection{Wideband RIS Designs}

Wideband operation is a key requirement for mm-Wave RIS implementations targeting practical 5G frequency bands such as n257 and n258, which together span from 24.25 to 29.5~GHz and correspond to nearly 20\% fractional bandwidth. This requirement becomes more critical for emerging 6G systems, where RIS-assisted links are expected to operate over wider bandwidths and higher frequency ranges to support enhanced data rates and sensing capabilities~\cite{Ruiqi2025APS}. However, most conventional RIS unit cells exhibit narrowband behavior because their reflection phase response is dominated by a single resonant mode~\cite{Gros2021mmWaveRIS},~\cite{Shekhawat2025mmWaveRIS}. As a result, the phase difference between discrete tuning states can only be preserved around the resonance frequency, while the reflection phase responses of different tuning states become highly dispersive with frequency outside this region, limiting array-level wideband beamforming performance in both 5G and future 6G scenarios.

From a hardware perspective, wideband RIS operation requires the unit cell to maintain a stable phase difference and a sufficiently high reflection magnitude over a broad frequency range, rather than being optimized at a single frequency point. To solve this challenge,~\cite{Wang_ruiqi2024TAP} proposes a wideband mm-Wave RIS design using a geometrically engineered patch-based unit cell that exploits multi-mode excitation within a simple structure. By appropriately shaping the patch geometry, two adjacent resonant modes are excited and the operating band is positioned between these resonances, thereby realizing broadband phase stability without increasing structural complexity.

Using this dual-mode excitation technique, the proposed unit cell achieves a phase difference of approximately $180^\circ$ with a tolerance of $\pm 20^\circ$ from 22.7 to 30.5~GHz, corresponding to a fractional bandwidth of 29.3\%. Meanwhile, the reflection magnitude of both tuning states remains higher than $-2.8$~dB across the same frequency range. The unit-cell size is approximately $0.35\lambda_0 \times 0.35\lambda_0$, which helps suppress grating lobes and improves angular stability at mm-Wave frequencies. Reconfigurability is realized using p-i-n diodes that control the unit cell between ON and OFF states. To minimize RF degradation, the switches and biasing network are placed on the backside of the ground plane and connected to the reflective elements through vias, with RF chokes used to isolate the dc bias from the RF signal path. This biasing strategy preserves the integrity of the reflective aperture while allowing dense routing suitable for large-scale RIS arrays.

\begin{figure}[t]
  \centering
  \subfloat[]{
    \includegraphics[width=0.99\linewidth]{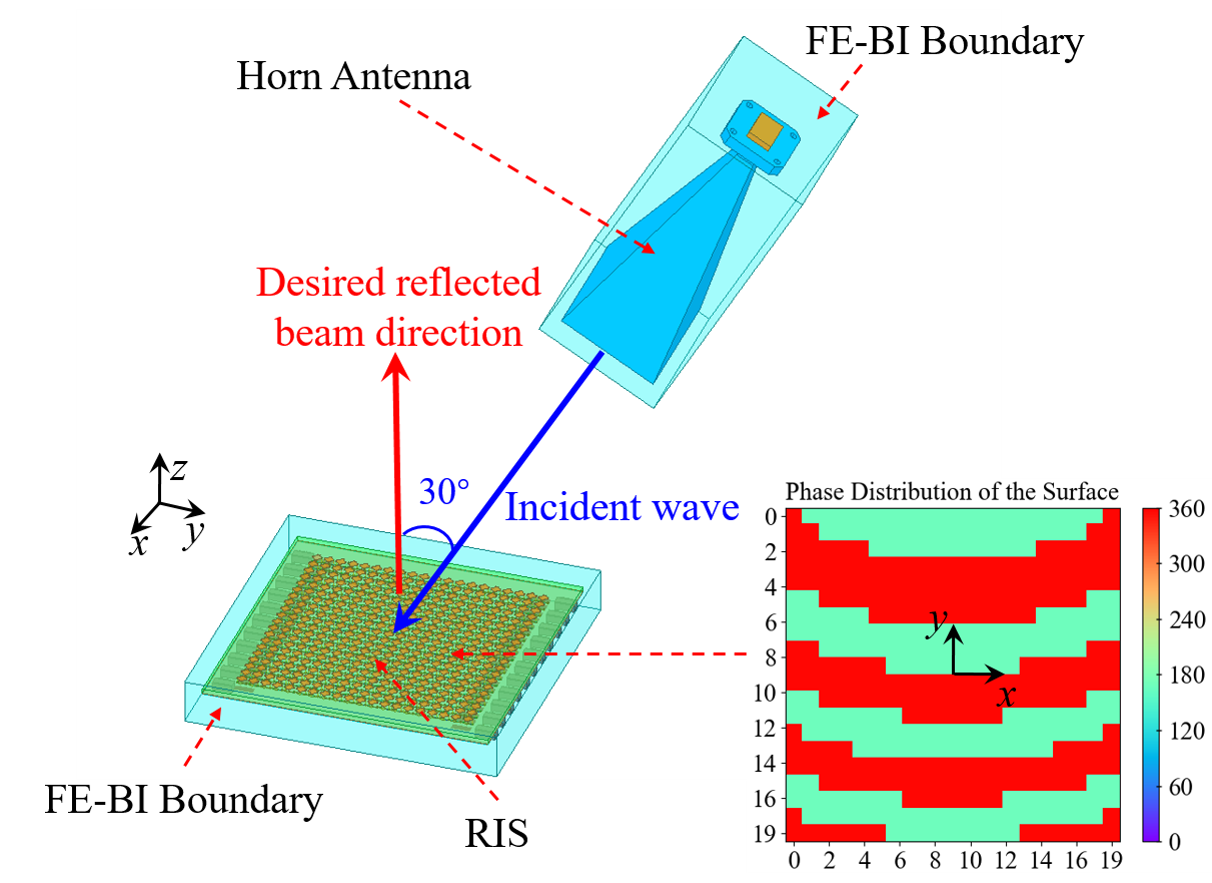}
    \label{fig:mmWave_RIS_simulation}
  }
  \vspace{0mm}
  \subfloat[]{
    \includegraphics[width=0.99\linewidth]{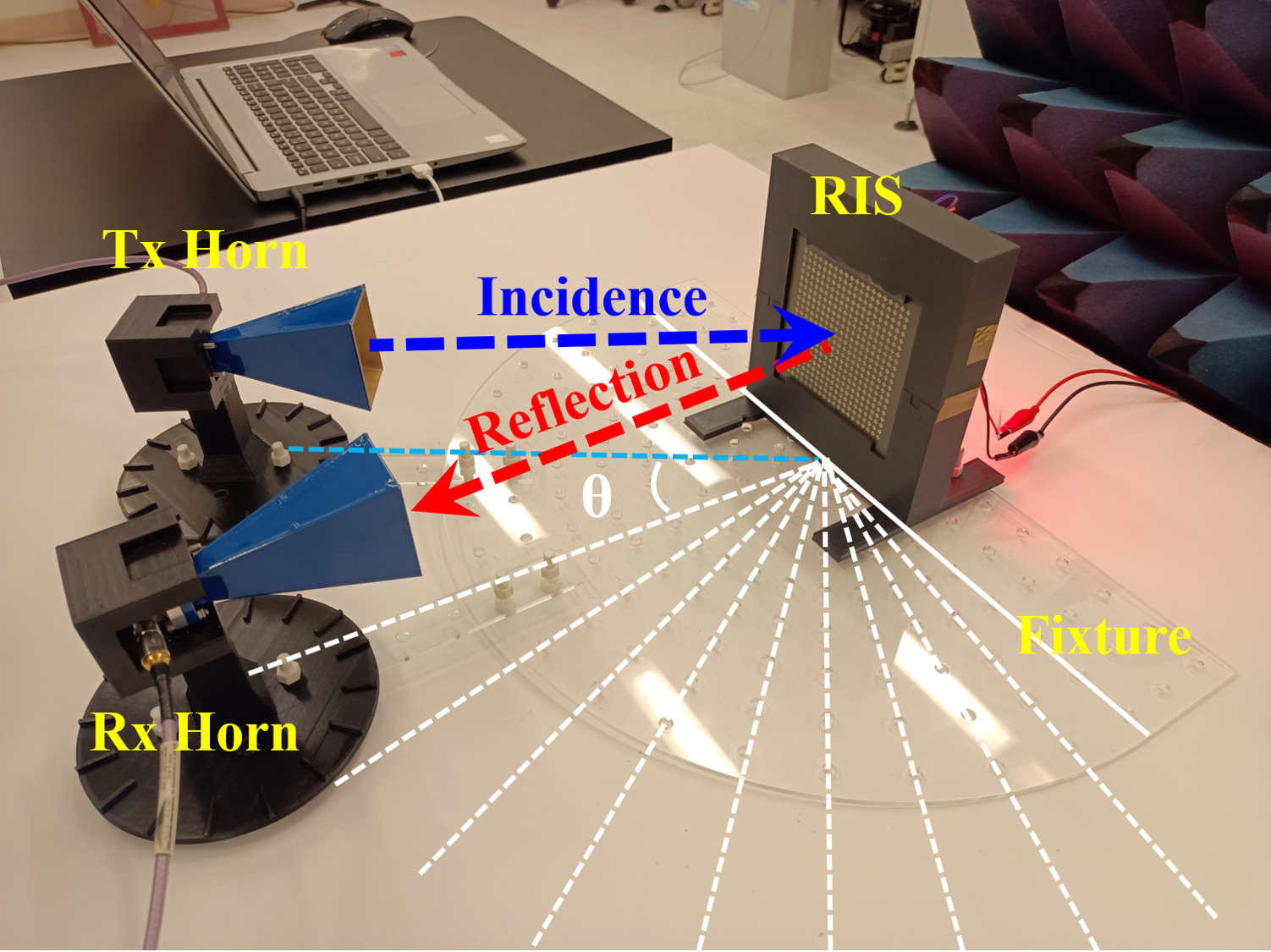}
    \label{fig:mmWave_RIS_prototype}
  }
  \caption{
    The wideband mm-Wave RIS hardware design in~\cite{Wang_ruiqi2024TAP}.
    (a) Full-wave simulation model.
    (b) Fabricated prototype and measurement.
  }
  \label{fig_mmWave_RIS}
\end{figure}

Based on the wideband unit-cell concept, a 20 × 20 RIS prototype is designed, fabricated, and experimentally characterized, as summarized in Fig.~\ref{fig_mmWave_RIS}. Fig.~\ref{fig_mmWave_RIS}(a) illustrates the full-wave simulation model and illumination configuration adopted to evaluate the wideband reflective behavior, while Fig.~\ref{fig_mmWave_RIS}(b) shows the fabricated prototype together with the measurement setup. Measurements demonstrate that the array maintains a 3-dB peak gain variation from 22.5 to 29.5 GHz, fully covering the n257 and n258 bands. Both near-field and far-field experiments confirm stable wideband beamforming behavior, validating the effectiveness of the proposed wideband design methodology from electromagnetic modeling to system-level implementation.

Overall, this work demonstrates that wideband mm-Wave RIS operation can be practically realized by exploiting geometry-induced dual-mode excitation within a simple 1-bit switching architecture, providing a scalable and hardware-efficient solution for broadband RIS-assisted mm-Wave systems. Building upon the proposed dual-mode resonance concept, subsequent works have further extended the operational bandwidth by increasing the number of resonant modes. By introducing multimode resonances, multiple phase transition points can be distributed across frequency, enabling the $180^\circ \pm 20^\circ$ phase range to be maintained over an ultrawide bandwidth spanning the n257–n261 mm-Wave bands, thereby achieving near-continuous spectral coverage~\cite{Hu2025mmWaveRIS}. This line of work illustrates a clear design evolution from dual-mode to multimode resonance engineering, establishing a general hardware methodology for wideband and ultrawideband mm-Wave RIS implementations.

\subsection{Phase Quantization and Its Practical Effects}

Due to hardware simplicity, most experimentally demonstrated wideband mm-Wave RIS designs adopt 1-bit phase quantization, where each unit cell provides only two discrete reflection phase states. Representative examples include the wideband RIS reported in~\cite{Wang_ruiqi2024TAP} and~\cite{Hu2025mmWaveRIS}, both of which successfully achieve stable wideband operation through geometry-engineered unit cells while relying on binary phase control. Such design choices are attractive for current 5G deployments and are also widely regarded as baseline solutions for early-stage 6G RIS prototyping. Although 1-bit phase quantization is effective in reducing circuit complexity, its coarse phase discretization inherently limits wavefront shaping accuracy. From an aperture perspective, the ideal continuous phase distribution required for precise beam steering in both 5G and future 6G systems is approximated by only two phase states, resulting in large phase errors distributed across the RIS surface. These phase errors manifest as elevated sidelobe levels in the near field and the emergence of quantization lobe in the far field. Such effects become increasingly pronounced at higher carrier frequencies, where electrically large apertures and narrow beamwidths make the radiation characteristics highly sensitive to phase inaccuracies.

\begin{figure}[t]
  \centering
  \subfloat[]{
    \includegraphics[width=0.9\linewidth]{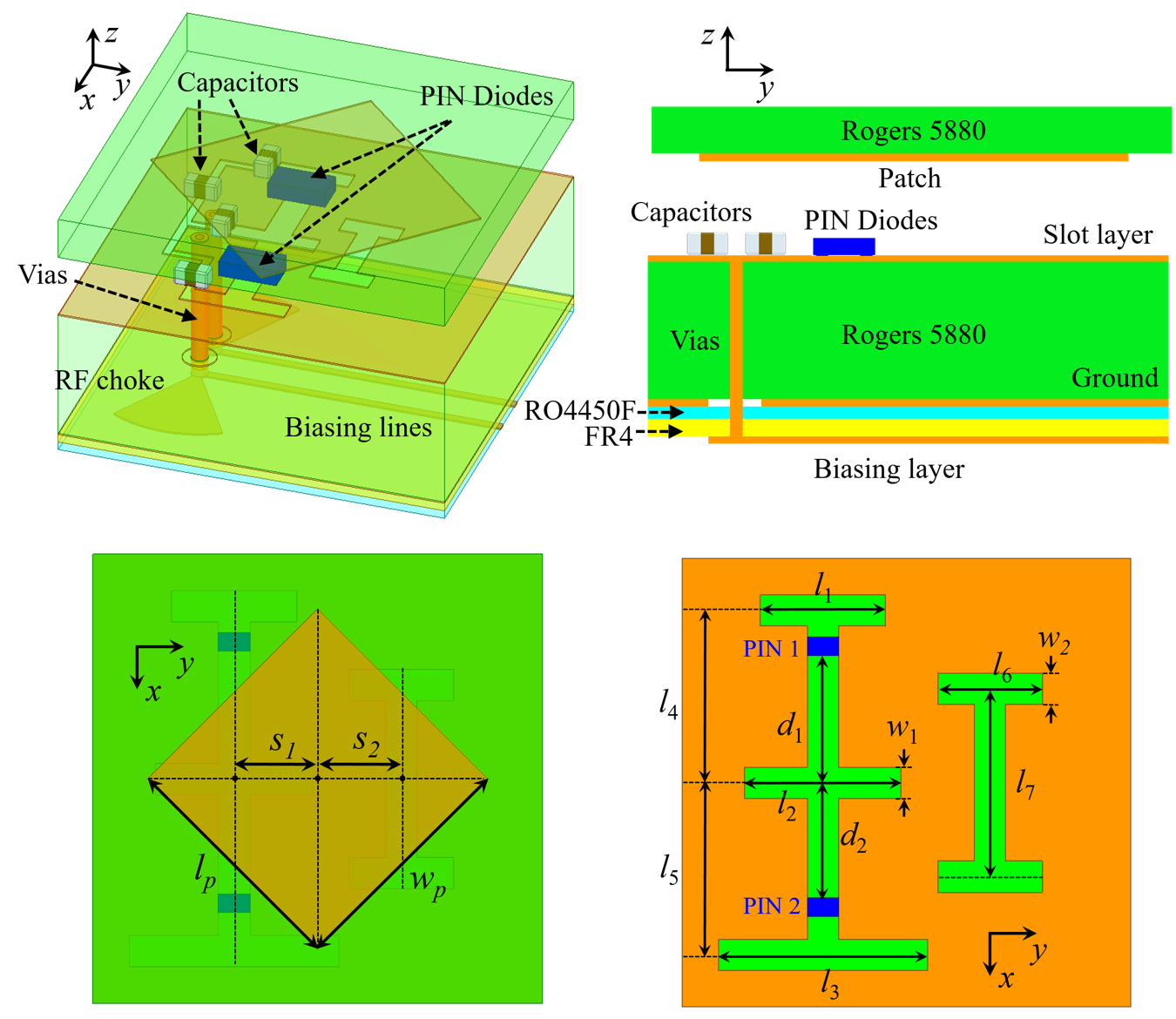}
    \label{fig:2_Bit_RIS_UnitCell}
  }
  \vspace{0mm}
  \subfloat[]{
    \includegraphics[width=0.9\linewidth]{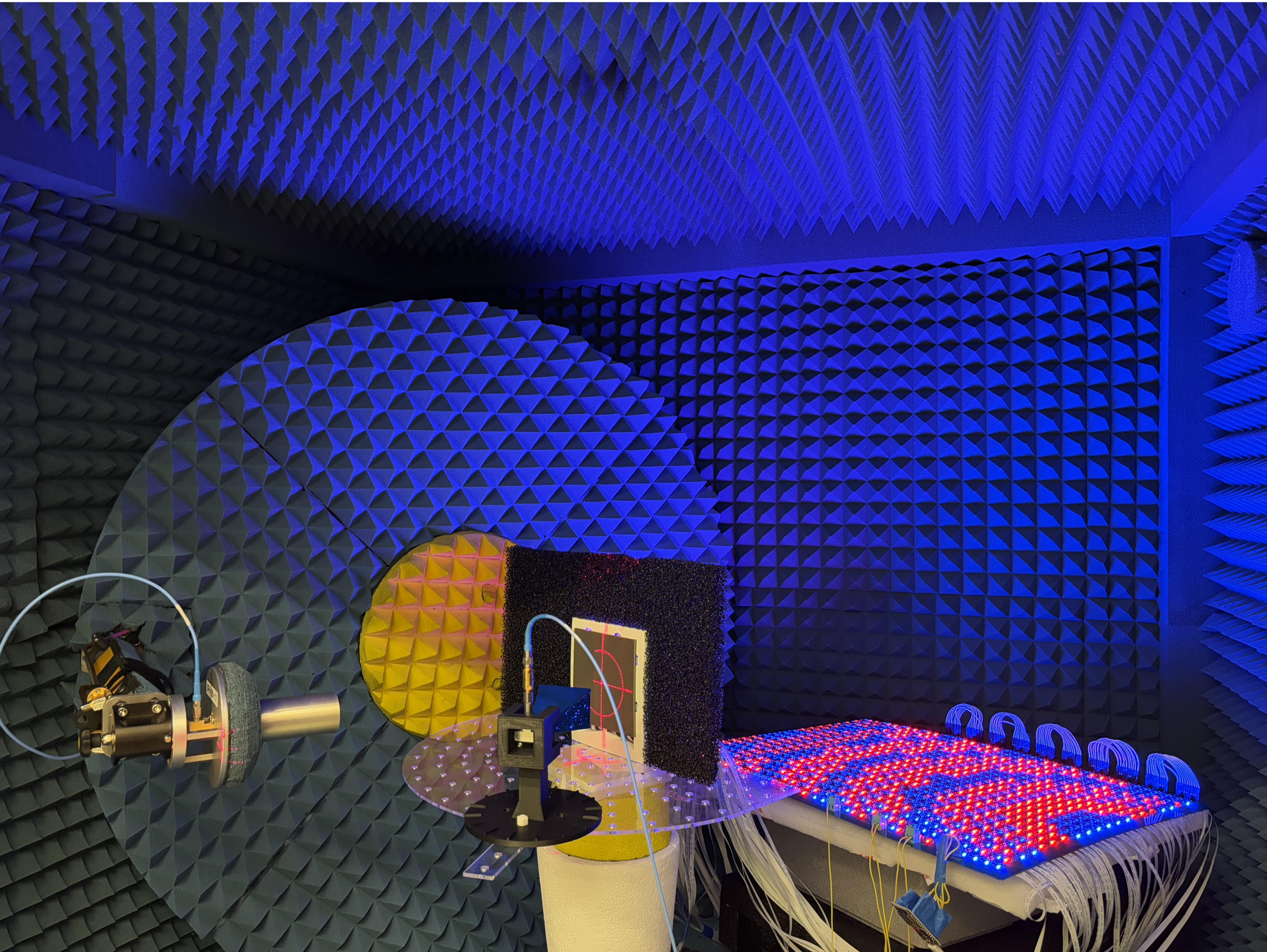}
    \label{fig:2_Bit_Measurement}
  }
  \caption{
    The 2-bit wideband mm-Wave RIS hardware design in~\cite{Ruiqi2025RIS}
    (a) Unit cell design.
    (b) Fabricated prototype and measurement.
  }
  \label{fig_2_Bit_RIS}
\end{figure}

While various mitigation strategies have been explored to suppress quantization lobes in 1-bit RIS designs, such as incorporating fixed random phase delays~\cite{Shekhawat2025QLL} and shifting half of the unit cells along the normal direction~\cite{Vabichevich2023QLL}, these methods do not fundamentally overcome the intrinsic phase resolution constraint imposed by binary quantization. These limitations motivate the consideration of multibit phase quantization as a practical means of improving beamforming quality while preserving hardware efficiency. Fig.~\ref{fig_2_Bit_RIS} illustrates the designed 2-bit wideband mm-Wave RIS design~\cite{Ruiqi2025RIS}, including the unit-cell configuration and the fabricated prototype for experimental validation. Compared to 1-bit designs, the 2-bit RIS provides four discrete phase states, which significantly reduce the phase error across the RIS aperture and allow the reflected wavefront to more closely approximate the ideal continuous-phase profile required for high-quality beamforming. Quantitative comparisons further highlight the practical benefits of increasing the phase resolution from 1-bit to 2-bit. Within the band of interest, the designed 2-bit RIS achieves a near-field sidelobe level of $-15.4\,\mathrm{dB}$, representing a $7.6\,\mathrm{dB}$ reduction compared to the corresponding 1-bit wideband RIS reported in~\cite{Wang_ruiqi2024TAP}. More importantly, in the far field, the quantization lobe level is suppressed to $-14.6\,\mathrm{dB}$ for the 2-bit RIS, whereas the 1-bit design exhibits a quantization lobe comparable to the main lobe. These improvements are directly relevant not only for interference-limited 6G systems, where stringent spatial selectivity and interference control are expected to be even more critical~\cite{Kim2025QLL}. Meanwhile, the proposed 2-bit RIS employs only two p-i-n diodes per unit cell and exploits multi-resonant behavior to maintain stable phase separation over a wide frequency range. This design enables wideband 2-bit operation while preserving a compact unit-cell size and a scalable biasing architecture, making it suitable for both current 5G and future 6G hardware platforms.


\subsection{Fully-Printed Low-Cost RIS}

While wideband RIS and multi-bit RIS designs have demonstrated significant performance improvements in terms of bandwidth, phase resolution, and beamforming accuracy, most of the reported implementations rely on conventional PCB-based fabrication processes. These designs typically employ discrete semiconductor devices such as p-i-n diodes, varactors, or MEMS switches, together with multilayer substrates and via interconnections. As a result, they inevitably suffer from high fabrication cost, complex assembly, and limited scalability when the number of unit cells increases to the order of hundreds or thousands. This cost and manufacturing bottleneck becomes even more critical for large-area RIS deployments envisioned for practical 5G and future 6G systems.

\begin{figure}[t]
  \centering
  \includegraphics[width=0.9\linewidth]{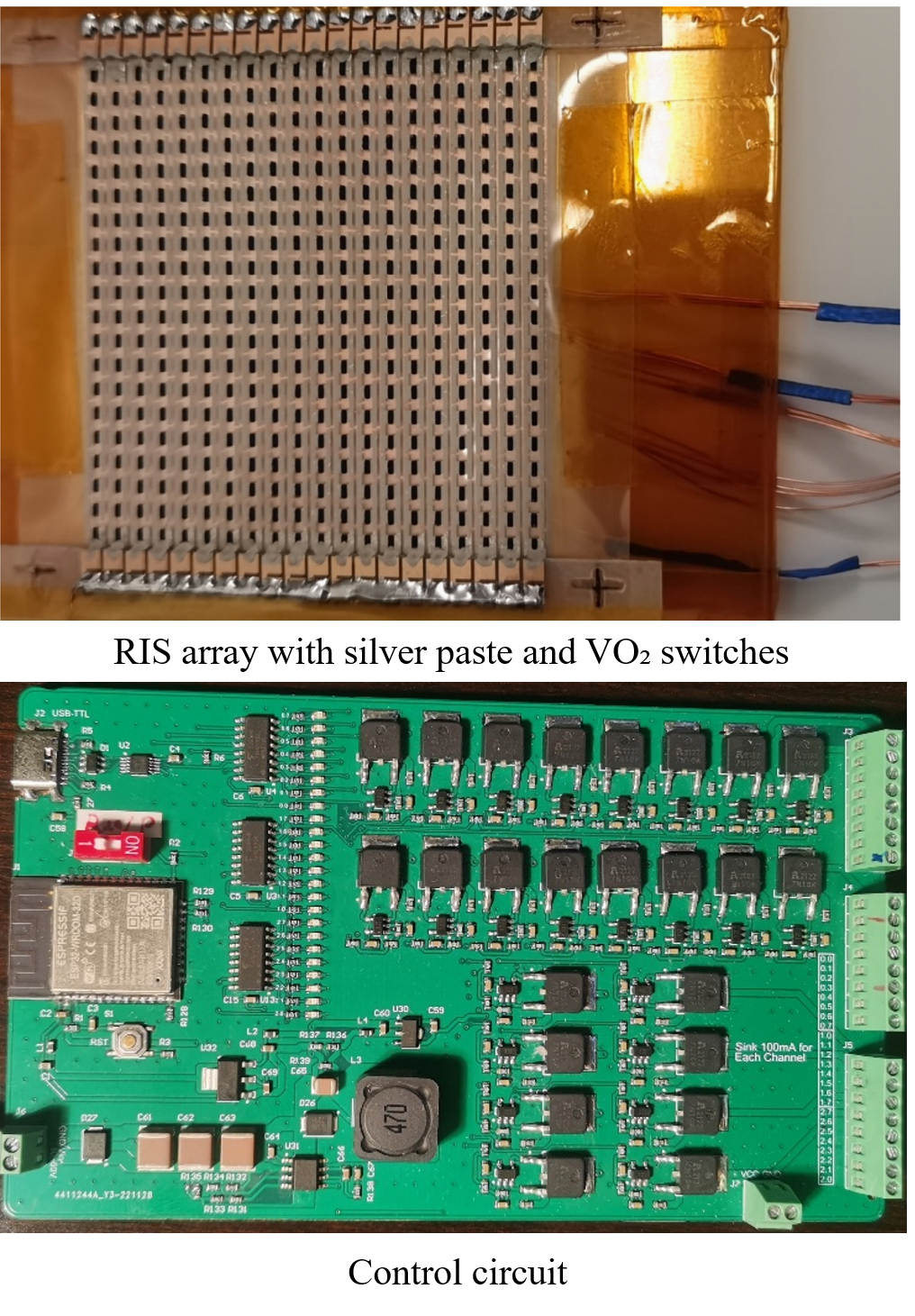}
  \caption{
    The fully-printed mm-Wave RIS hardware prototype with control circuit in~\cite{Yiming2025fullyprinted}
  }
  \label{fig_fully-printed_RIS}
\end{figure}

To address these limitations, additive manufacturing technique has recently attracted increasing attention~\cite{Ruiqi2023RFID},~\cite{Ruiqi2023DSSRR},~\cite{gibson2021additive}, ~\cite{Ruiqi2023Isotropy}. In particular, screen printing~\cite{Ruiqi2021Asymmetric} enables high-throughput, large-area, and low-cost fabrication, making it a promising candidate for realizing scalable RIS hardware. Fig.~\ref{fig_fully-printed_RIS} illustrates the first fully screen-printed RIS prototype reported in ~\cite{Yiming2025fullyprinted} that replaces conventional PCB processes and discrete semiconductor switches with printed conductors and printed phase-change-material-based switches.

In this design, both the metallic resonators and the biasing networks are realized using silver paste printed directly on a flexible substrate~\cite{Yiming2023APS}, while the switching functionality is achieved using screen-printed vanadium dioxide (VO$_2$) switches. Unlike p-i-n diodes or varactors, VO$_2$ exhibits an intrinsic metal--insulator transition that can be triggered by electrical current or thermal excitation, enabling reconfigurable EM responses without the need for discrete soldered components. The RIS array adopts a via-less and single-layer topology, which is inherently compatible with the screen-printing process and avoids alignment and reliability issues associated with multilayer PCB stacking. A key challenge in fully printed RIS implementations lies in maintaining wideband performance while minimizing conductor and switch losses. By carefully co-designing the resonator geometry and the printed VO$_2$ switch dimensions, the presented RIS achieves a stable phase difference close to $180^\circ$ over a wide frequency range covering the 5G n257 and n258 bands.

The notable advantage of the fully printed RIS is its extremely low fabrication cost. Since both the conductors and switches are deposited using screen printing, the cost is dominated by raw materials rather than discrete RF components or precision PCB processing. For a typical $20\times20$ RIS array, the total switch cost can be reduced to only a few US dollars, which is orders of magnitude lower than that of PCB-based RIS implementations employing commercial semiconductor switches. In addition, the solder-free fabrication process significantly improves manufacturing yield and enables flexible and conformal implementations.

\subsection{Optically Transparent RIS}

In many practical deployment scenarios, RIS are expected to be integrated into existing urban infrastructures such as building windows, glass façades, vehicle windshields, display panels, and other visually sensitive surfaces, where optical transparency becomes a critical requirement. Conventional PCB-based RIS architectures, although mature and efficient from an EM perspective, are inherently opaque and therefore incompatible with such environments. This limitation has motivated growing research interest in optically transparent RIS, which aim to manipulate EM waves while allowing visible light to pass through with minimal visual obstruction.

\begin{figure}[t]
  \centering
  \includegraphics[width=0.9\linewidth]{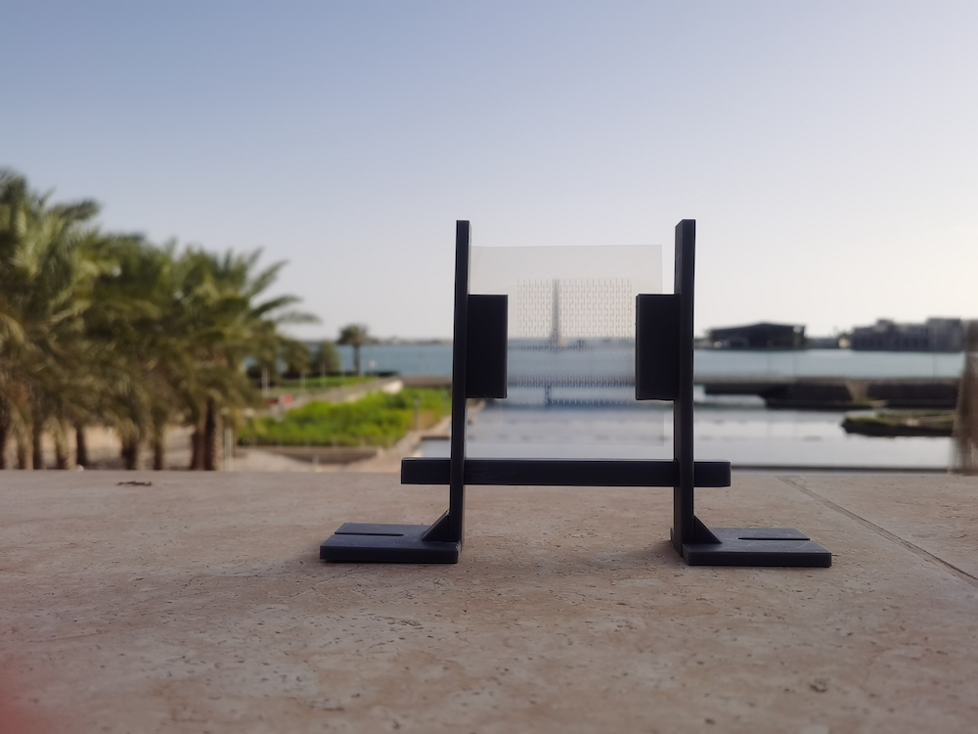}
  \caption{
    The mm-Wave static optically transparent RIS prototype in~\cite{Yiming2025Transparent}
  }
  \label{fig_Optically_Transparent_RIS}
\end{figure}

From a hardware perspective, optically transparent RIS realization depends on the choice of conductive materials and substrates. Metallic mesh structures achieve transparency by reducing metal coverage, but their required feature size at mm-Wave frequencies becomes extremely small, leading to fabrication complexity and increased conductor loss~\cite{liang2024optically}. Transparent conductive oxides such as indium tin oxide offer good optical uniformity but suffer from low electrical conductivity and high cost, which degrade reflection efficiency at mm-Wave bands. Conductive polymers provide flexibility and transparency, yet their limited conductivity prevents broadband mm-Wave RIS operation. To overcome these limitations, conductive nanomaterials, particularly silver nanowires, emerge as a promising solution. Silver nanowire films form nanoscale conductive networks that enable a favorable trade-off between optical transparency and electrical conductivity, allowing visible light transmission while maintaining sufficient RF conductivity at mm-Wave frequencies. Moreover, silver nanowire inks support additive manufacturing techniques such as screen printing, enabling low-cost and scalable large-area fabrication.

A representative optically transparent mm-Wave RIS implementation based on screen-printed silver nanowires is reported in \cite{Yiming2025Transparent}. In this work, the RIS unit cells are realized on transparent substrates using silver-nanowire-based conductive patterns combined with discrete p-i-n diodes for reconfigurability. The via-less architecture and additive printing process enable optical transparency exceeding 80 percent while maintaining broadband operation across the 5G n257 and n258 bands. Experimental results demonstrate that the proposed transparent RIS achieves 10--15~dB gain enhancement over a wide frequency range with stable performance under oblique incidence, validating its practical feasibility for real-world mm-Wave deployments.

Compared with metal-mesh- and indium-tin-oxide-based designs, silver-nanowire-based transparent RIS offer several distinct advantages. First, the nanoscale feature size of the nanowire network mitigates the scalability issues encountered by metallic meshes at high frequencies. Second, the screen-printing process significantly reduces fabrication cost and enables rapid prototyping over large areas, which is essential for RIS panels consisting of hundreds or thousands of unit cells. Third, the mechanical flexibility of silver nanowire films allows conformal integration onto curved or non-planar transparent surfaces, further expanding the application space of RIS-enabled smart environments. Despite these advantages, optically transparent RIS still face several open challenges. The conductivity of silver nanowire films remains lower than that of bulk metals, leading to increased ohmic loss compared with opaque PCB-based RIS. In addition, the integration of biasing networks and discrete switching components must be carefully designed to preserve both optical transparency and EM performance. Long-term reliability under environmental exposure, including humidity, temperature variation, and mechanical stress, also requires further investigation for large-scale deployment.


\subsection{From 2D RIS to 3D Architectures}

Most existing RIS are implemented as planar two-dimensional structures, where EM wave manipulation is confined to a single physical surface. While such 2D RIS designs have demonstrated effective beam steering and channel enhancement in many scenarios, their inherent planar geometry fundamentally restricts spatial coverage to one hemisphere. As a result, blind regions inevitably arise, particularly for incident or desired propagation directions near endfire or outside the illuminated half-space. This limitation becomes increasingly pronounced at mm-Wave frequencies, where propagation is highly directional and susceptible to blockage. To extend the functionality, simultaneously transmitting and reflecting RIS (STAR-RIS) have been proposed to enable signal manipulation on both sides of a planar surface, improving spatial utilization over purely reflective designs. However, their planar geometry still fundamentally limits spatial coverage, leaving gaps in three-dimensional deployments, particularly near endfire directions~\cite{Xiang2023STAR}.

To overcome the intrinsic spatial limitations of planar RIS, recent research~\cite{Ruiqi2026RIS} has begun to explore three-dimensional RIS architectures, where multiple RIS surfaces are arranged in space to enable volumetric EM wave control. By extending RIS from a single surface to a spatial assembly of surfaces, 3D architectures provide an additional degree of freedom for redirecting incident waves toward neighboring directions that are inaccessible to planar designs. Such configurations are particularly attractive for mm-Wave systems deployed in complex environments, where signals may arrive from or need to be redirected toward multiple spatial directions due to dynamic blockage and mobility.

\begin{figure}[t]
  \centering
  \includegraphics[width=1\linewidth]{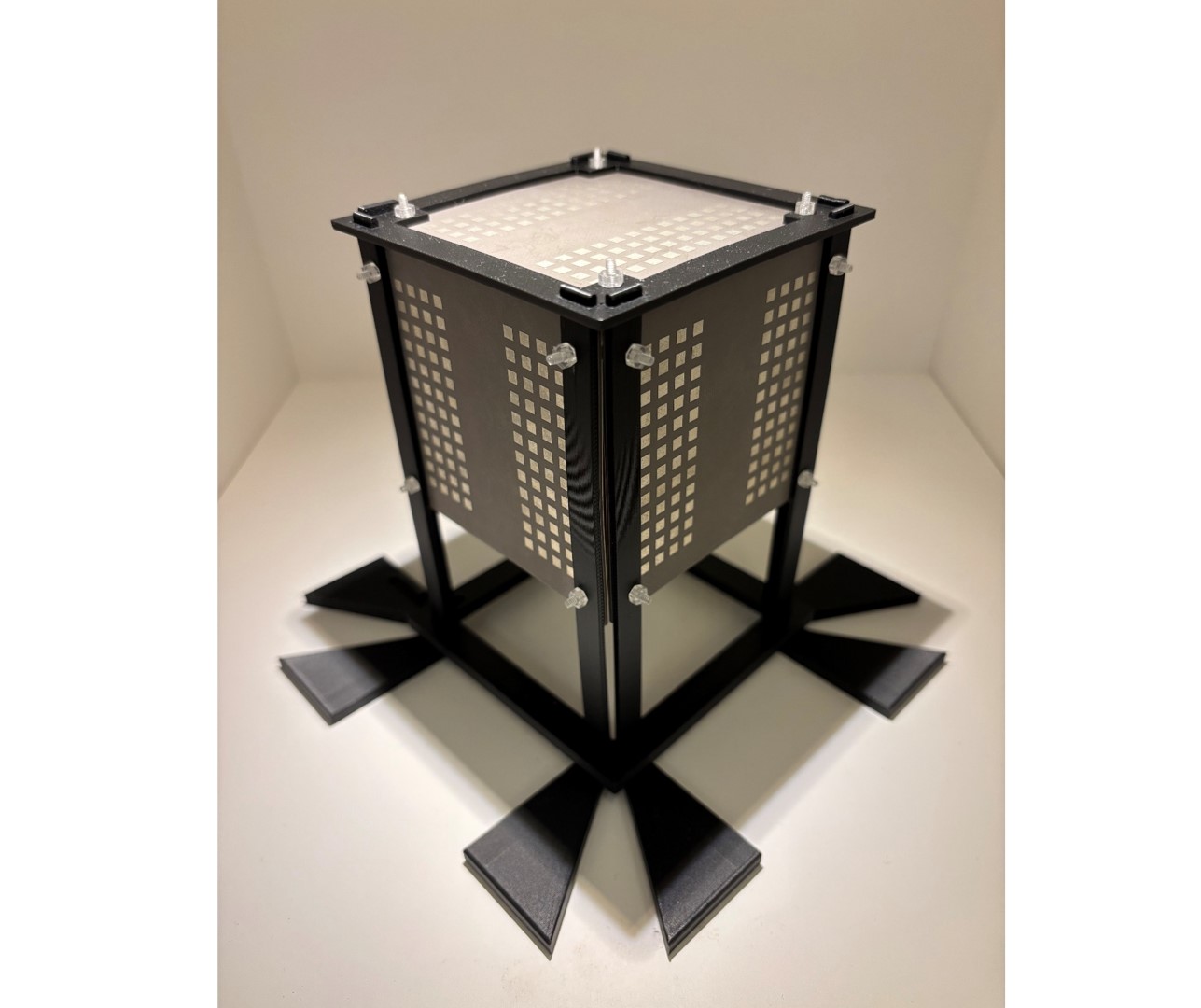}
  \caption{
    The mm-Wave 3-D RIS prototype in~\cite{Ruiqi2026RIS}
  }
  \label{fig_3D_RIS}
\end{figure}

A representative realization of this concept is a cube-based 3D RIS architecture, as demonstrated in~\ref{fig_3D_RIS} where multiple planar RIS panels are assembled to form a volumetric structure. In this configuration, EM waves can be not only reflected by the illuminated surface but also guided toward adjacent surfaces through controlled inter-surface routing. This mechanism enables spatial beam redirection across different faces of the structure, effectively mitigating blind regions inherent to planar RIS designs. Importantly, this volumetric functionality is achieved without fundamentally altering the underlying planar RIS unit-cell design, which preserves fabrication maturity and scalability.

From a practical hardware perspective, the effectiveness of such a 3D RIS architecture has been experimentally validated at mm-Wave frequencies. In the implemented prototype, measured results demonstrate that a single-surface reflection configuration achieves gain enhancements of up to 14.7 dB at 26 GHz compared to the RIS OFF state. More importantly, when the incident wave is redirected toward a neighboring surface through the 3D architecture, a measured gain enhancement of 14.1 dB is achieved. These results experimentally confirm that volumetric wave manipulation enables effective signal redirection beyond the illuminated surface, which is fundamentally unattainable using conventional planar RIS.


At the communication-system level, the volumetric advantages of the proposed 3D RIS architecture are validated through over-the-air experiments using a mm-Wave software-defined radio platform. When the RIS is activated, clear improvements in constellation quality and error vector magnitude (EVM) are observed. More importantly, these results demonstrate that the additional spatial degrees of freedom enabled by the 3D RIS architecture translate into tangible communication-system benefits, confirming that volumetric wave manipulation can provide performance enhancements beyond conventional planar implementations.


\subsection{Wideband RIS with Tunable Phase-Frequency Profiles}

In wideband communication systems, the reflection response of practical RIS unit cells is inherently frequency dependent due to their resonant nature, rendering the conventional assumption of frequency-flat or independently configurable phase shifts across subcarriers physically unrealistic. This limitation motivates the concept of RISs with \emph{tunable phase-frequency profiles}, where each unit cell is characterized not only by its reflection phase at a reference frequency but also by the slope of its phase response over the operating band.

From a system-level perspective, modeling each RIS element via a resonant equivalent circuit leads to a parametric phase-frequency response that can be compactly described by a phase intercept and a controllable slope. By discretizing these parameters, an RIS can switch among a finite set of realizable phase-frequency profiles. Joint optimization of the selected profiles across RIS elements and the power allocation over OFDM subcarriers enables coherent signal combining over frequency, effectively compensating for geometry-dependent propagation delays. Such slope-aware RIS designs have been shown to significantly improve the achievable rate in wideband OFDM systems compared to conventional RIS architectures with fixed phase-frequency slopes, even in the presence of moderate reflection loss~\cite{AbbasTWC2025}.

From a hardware realization standpoint, independent control of the phase center and phase-frequency slope can be achieved using multilayer unit-cell architectures composed of resonant elements with distinct inductive and capacitive characteristics. In particular, cascaded dog-bone and patch resonators loaded with varactors enable continuous and decoupled tuning of these two parameters by adjusting the bias voltages. Experimental results confirm that such designs can span a wide range of phase-frequency slopes while maintaining low reflection loss, thereby providing a practical electromagnetic foundation for slope-adaptive RISs in wideband systems~\cite{AbbasTAP2025}. Together, these advances establish tunable phase-frequency profiles as a key enabler for bridging electromagnetic constraints and system-level optimization in next-generation RIS-assisted wideband communications.

\subsection{Liquid-Crystal-Based RIS}

Liquid-crystal (LC)-based RIS provides continuous phase tunability with voltage-driven operation and negligible steady-state current, making it attractive for mm-Wave beam-steering systems~\cite{Kim2024mmWaveRIS}. By exploiting the anisotropic permittivity of LC under an applied electric field, the reflection phase can be smoothly tuned without discrete quantization states, enabling analog phase control over wide frequency ranges.

\begin{figure}[t]
  \centering
  \includegraphics[width=0.9\linewidth]{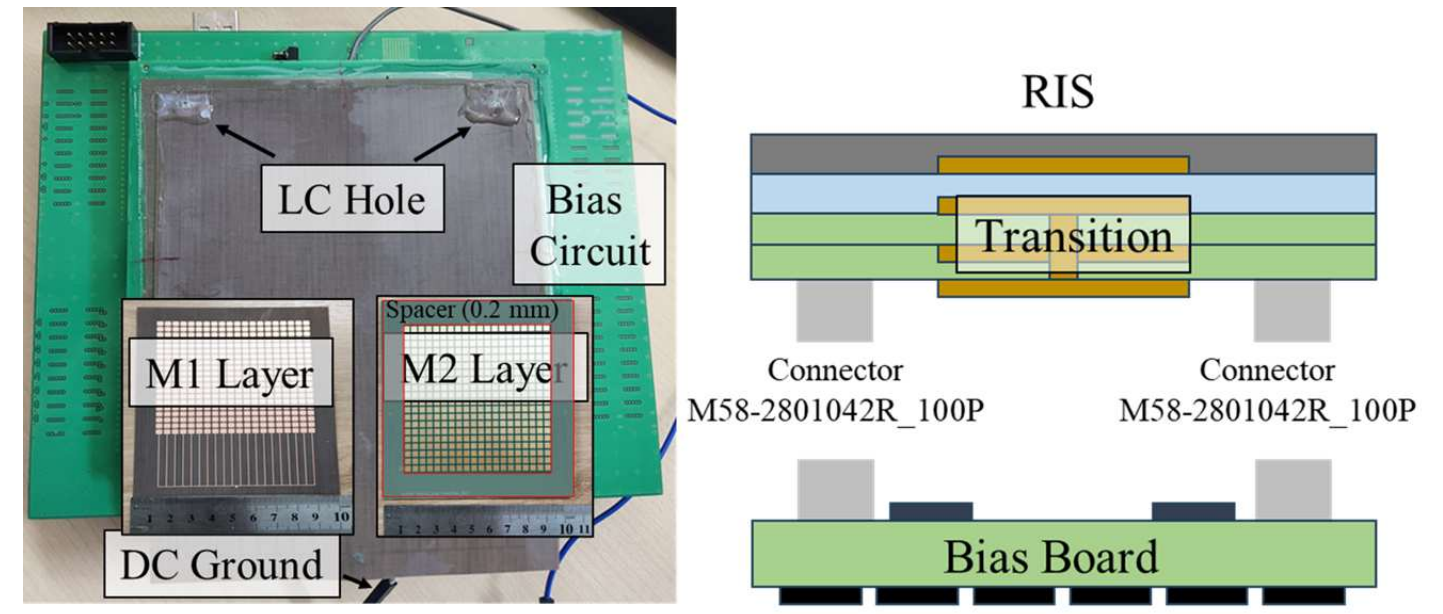}
  \caption{
    The mm-Wave LC-based RIS prototype in~\cite{kim2026lowpower}
  }
  \label{fig_LC_RIS}
\end{figure}

A representative 28~GHz LC-based RIS is reported in~\cite{kim2026lowpower}, where a multilayer unit cell incorporating an LC cavity achieves over 200$^\circ$ phase tuning across 27--30~GHz, as shown in Fig.~\ref{fig_LC_RIS}. An active addressing scheme reduces the bias network from $N^2$ to $2N$ nodes for an $N \times N$ array, improving scalability for electrically large apertures. Furthermore, pulse-based biasing with storage capacitors enables beam-sustainable operation, significantly lowering average power consumption.

Despite these advantages, LC-based RIS is mainly limited by molecular relaxation dynamics, resulting in slower reconfiguration speeds compared with semiconductor-based implementations. Therefore, LC-RIS is particularly suitable for low-power, quasi-static mm-Wave scenarios, such as fixed wireless access and infrastructure-assisted coverage enhancement, where energy efficiency and large-aperture scalability are prioritized over ultra-fast switching.

\subsection{On-Chip RIS Designs for 6G}

While early RIS prototypes were realized using PCB or printing technologies, the extension of RIS toward sub-THz 6G bands changes the hardware constraints. At frequencies around 100 GHz and above, the unit-cell dimensions shrink to the order of hundreds of micrometers, making interconnect parasitics and packaging loss comparable to the electromagnetic response of the RIS itself. This motivates a transition from board-level RIS toward on-chip RIS architectures that can be co-fabricated and co-packaged with transceivers. By leveraging CMOS-compatible platform, on-chip RIS designs enable fine-grained control of EM propagation on a sub-wavelength scale. Such architectures support desired phase tuning, compact unit-cell footprints well below $\lambda_0/2$, and dense two-dimensional integration with biasing and control circuitry. As a result, the RIS becomes scalable to large arrays and amenable to monolithic or heterogeneous integration with mm-wave and sub-THz transceivers, which is essential for short-range communication, imaging, and sensing in future 6G systems.

\begin{figure}[t]
  \centering
  \includegraphics[width=0.99\linewidth]{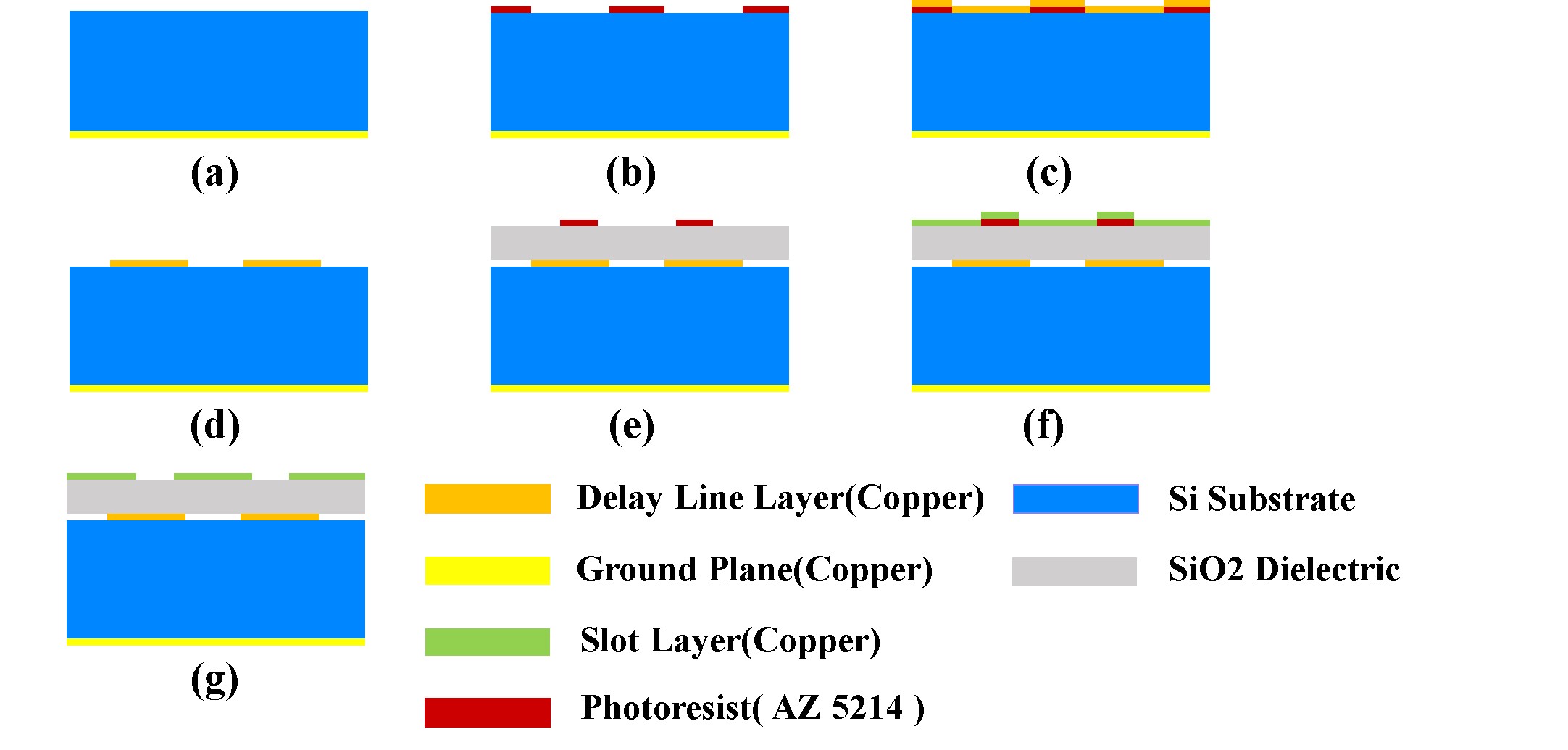}
  \caption{
    Fabrication process of the On-chip RIS. (a) Prepared 300-um-thick
high-resistivity silicon wafer with one copper layer sputtered on one side. (b)Lithography. (c)Sputtering of copper. (d)Lift-off. (e)-(g) Deposited SiO2 and repeated the processes of (b)-(d).
  }
  \label{fig_3_G_1}
\end{figure}

\begin{figure}[t]
  \centering
  \includegraphics[width=0.99\linewidth]{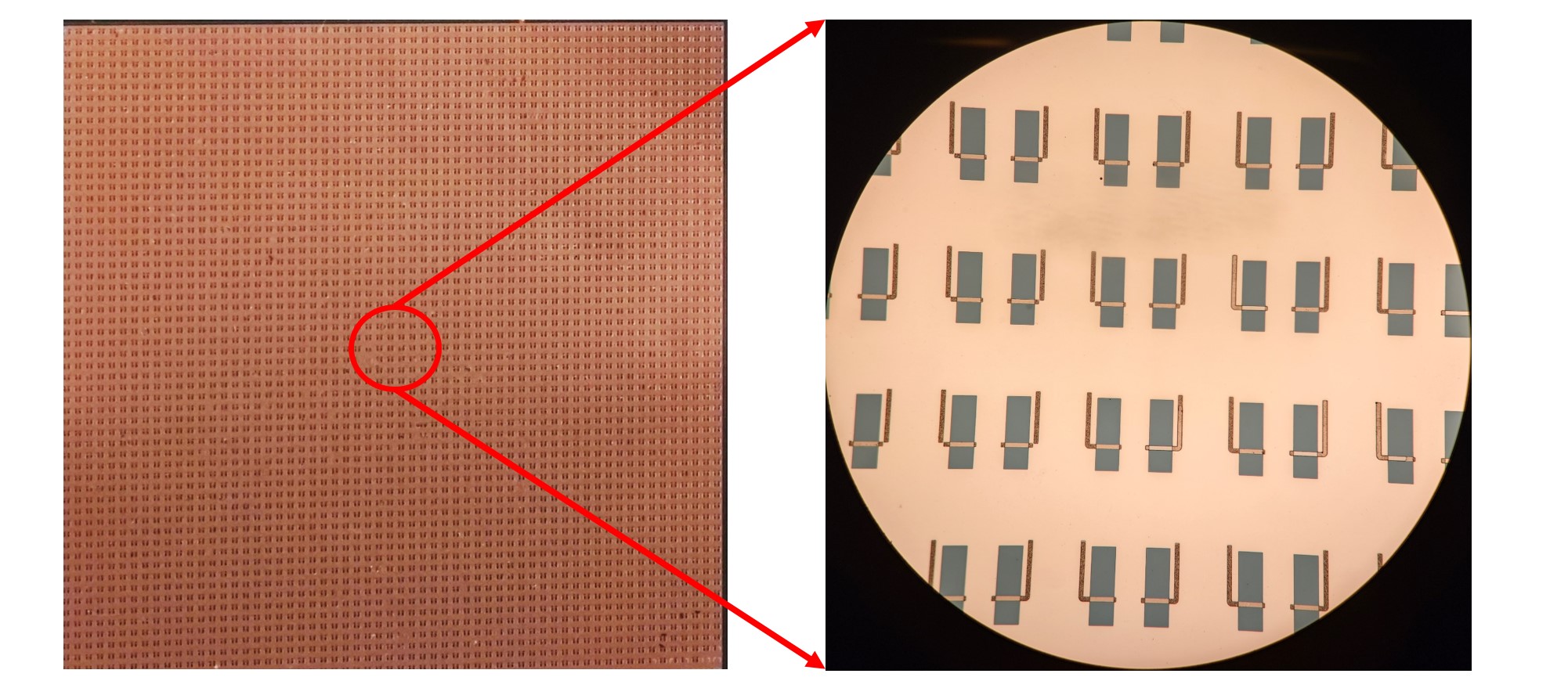}
  \caption{
    Photograph of the fabricated RIS-on-chip array and microscopic view of the unit cells in~\cite{su2025100ghzcmoscompatiblerisonchipbased}
  }
  \label{fig_3_G_2}
\end{figure}

Recent on-chip RIS implementations have adopted slot-coupled, phase-delay-line-based unit-cell structures, as reported in~\cite{su2025100ghzcmoscompatiblerisonchipbased}. 
In this design, a subwavelength slot in the top metal layer couples the incident wave into an on-chip phase-delay-line. A thin SiO$_2$ layer provides electrical isolation while preserving strong EM coupling. The effective electrical length of the transmission paths is controlled by integrated VO$_2$ phase-change sections, enabling two distinct reflection states with an approximately $180^\circ$ phase difference at 100.75~GHz and a reflection magnitude better than -1.2~dB. The compact unit-cell footprint of $0.23\lambda_0 \times 0.23\lambda_0$ further supports dense on-chip array integration. Following the unit-cell design, a large-scale RIS-on-chip array was realized using a CMOS-compatible thin-film process, as shown in~\ref{fig_3_G_1}. 
The fabrication begins with the deposition of a copper ground plane on a high-resistivity silicon substrate, followed by a patterned copper layer forming the phase-delay lines and a PECVD SiO$_2$ layer for insulation. The top metallic slot layer is then defined by photolithography, metal deposition, and lift-off to complete the multilayer RIS stack. The entire process is based on standard thin-film and lithographic steps, demonstrating full compatibility with semiconductor manufacturing technologies. 

Upon completion of the fabrication process, a RIS-on-chip array was obtained, as shown in~\ref{fig_3_G_2}. The left image presents the full $60 \times 60$ array on the silicon die, while the right image provides a magnified view of several unit cells. The experimental characterization of the RIS-on-chip array was performed in anechoic chamber. 
The measured reflection responses for the two reconfigurable states exhibit a clear and pronounced contrast, with a peak reflection enhancement of 27.1~dB observed in the designed steering direction. These results confirm that the fabricated chip-scale RIS operates as an electrically programmable reflective aperture at around 100~GHz, validating the CMOS-compatible RIS-on-chip implementation.

\red{\section{Hardware–System Considerations in Practical RIS Deployment}}

This section discusses critical issues in practical RIS implementations, with references to state-of-the-art studies.

\subsection{Mutual Coupling in RIS}

\begin{figure}[t]
  \centering
  \includegraphics[width=0.9\linewidth]{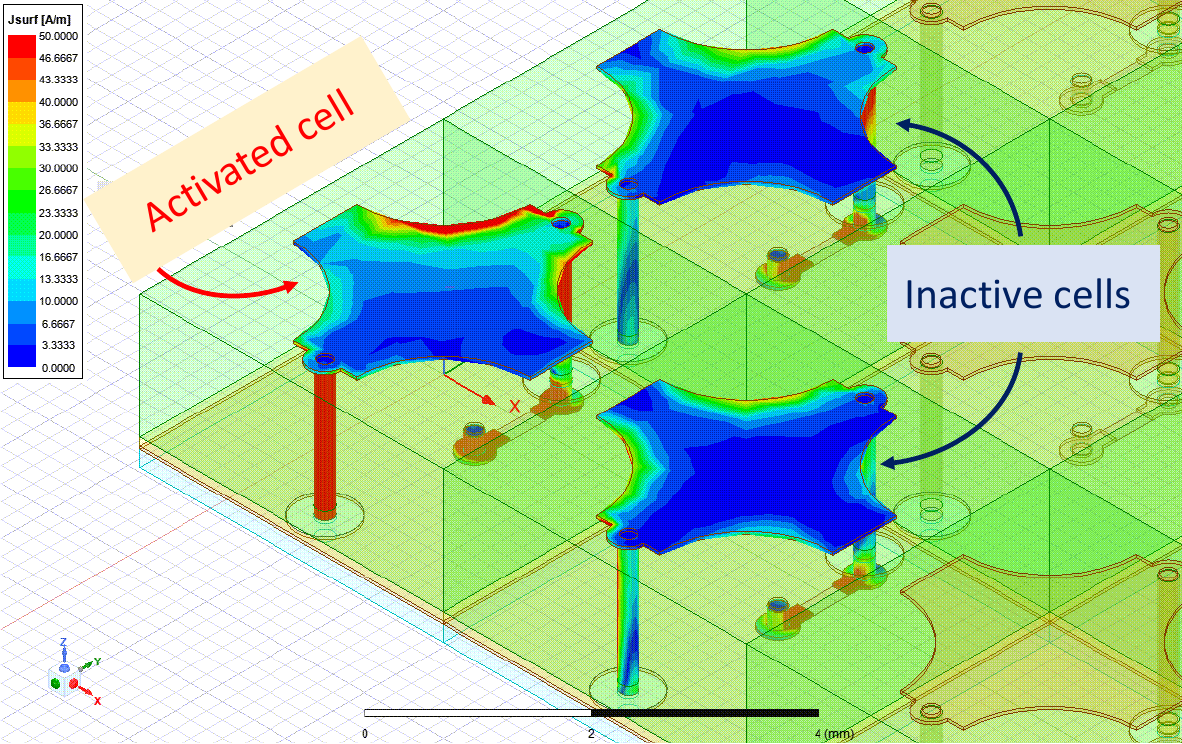}
  \caption{
    Surface current distribution illustrating the mutual coupling effect when a single unit cell is activated.
  }
  \label{fig_MC}
\end{figure}

When RIS unit cells are densely integrated, \emph{mutual coupling} becomes a significant practical concern. Mutual coupling refers to the phenomenon whereby the current excited at one reflective unit cell induces parasitic currents in its neighboring cells. As illustrated in Fig.~\ref{fig_MC}, even when only a single unit cell located at the corner is activated, noticeable surface currents can be observed on adjacent unit cells due to mutual coupling. These induced currents generate unintended radiation, which in turn distorts the desired reflection beam pattern.

When this effect is taken into account, the RIS response matrix $\Gammam$ in~\eqref{eq:yfreq} is no longer diagonal and is instead given by $(\Gammam^{-1}-\Sm)^{-1}$, where $\Sm$ denotes the scattering matrix among the RIS unit cells and characterizes the mutual coupling effect~\cite{Zheng2024Mutual}. Specifically, the $(i,j)$-th entry $[\Sm]_{i,j}$ represents the scattering parameter between the $i^\text{th}$ and $j^\text{th}$ unit cells, defined as~\cite{Pozar2011Microwave}
\begin{equation}
[\Sm]_{i,j} = \frac{V_i^{\mathrm{out}}}{V_j^{\mathrm{in}}}\bigg|_{V_k^{\mathrm{in}}=0,\ \forall k\neq j},
\end{equation}
where $V_j^{\mathrm{in}}$ denotes the incident voltage wave applied to the $j^\text{th}$ unit cell and $V_i^{\mathrm{out}}$ denotes the outgoing voltage wave at the $i^\text{th}$ unit cell, with all other unit cells terminated. An experimental validation of this model has been reported in~\cite{Zheng2024Mutual}.

In general, the impact of mutual coupling becomes pronounced when the spacing between RIS unit cells is smaller than half a wavelength. In such tightly packed arrays, neglecting mutual coupling can lead to substantial performance degradation in channel estimation and beamforming~\cite{Qian2021Mutual,Zheng2024Mutual2}. To address this issue, several mutual coupling-aware methods have recently been proposed for applications including channel estimation, beamforming, and localization~\cite{Fadakar2025Mutual,Del2026Ambiguity}. Interestingly, recent studies have shown that, when properly exploited, mutual coupling in RISs can even be leveraged to enhance system performance~\cite{Zheng2026Mutual,Nerini2026Global}.

\subsection{Calibration Considerations}

Accurate calibration is essential for RIS-aided wireless systems, as imperfect knowledge of RIS geometry and hardware states leads to model mismatches that degrade sensing, localization, and communication performance. In practice, these mismatches cannot be mitigated by simply increasing transmit power and instead impose fundamental performance limits~\cite{Ghazalian2025Calibration}.

RIS calibration can be broadly categorized into \emph{geometry calibration} and \emph{hardware calibration}. Geometry calibration estimates the RIS position and orientation, which is critical when RISs serve as reference anchors for localization and sensing and for enabling location-aware beamforming~\cite{Zheng2023Misspecified}. Hardware calibration compensates for nonideal RIS responses, including mutual coupling~\cite{Zheng2024Mutual2}, faulty unit cells~\cite{ozturk2024ris}, and phase-dependent amplitude variations~\cite{Liu2024Passive}, which distort the RIS radiation characteristics.

Calibration errors have a more severe impact on localization and sensing than on communication. While communication performance is mainly affected through SNR loss, localization accuracy depends critically on precise geometric and phase information and becomes mismatch-limited at high SNRs. Consequently, effective calibration, either agent-based~\cite{Zhan2026Estimating} or jointly performed with user localization~\cite{Zheng2024JrCUP}, is indispensable for fully exploiting RIS-aided wireless systems.

\subsection{Interference in Multi-RIS Systems}

As multi-RIS systems emerge as a practical means to enhance coverage and controllability in large-scale wireless networks~\cite{Katsanos2024Multi}, interference becomes a critical implementation challenge. Deploying multiple distributed RISs can strengthen the desired signal by creating additional propagation paths and improving spatial diversity, often outperforming a single centralized RIS with the same total number of elements. However, increasing the number of RISs simultaneously amplifies RIS-assisted inter-user interference. When the RIS density becomes excessive, particularly relative to the number of access points, the interference growth outweighs the signal enhancement, causing interference-limited performance and eventual throughput degradation~\cite{Nahhas2024Performance}. This behavior indicates the existence of an optimal number of RISs and highlights that interference-aware deployment and phase-shift optimization are essential for practical multi-RIS system design.

\subsection{Considerations in RIS Partitioning and Multi-Functional Operation}

RIS partitioning has emerged as a mechanism to enable task-oriented and multi-functional operation on a single RIS. However, from a practical deployment perspective, partitioning inherently introduces nontrivial design trade-offs. By virtually dividing the RIS aperture into multiple partitions, the effective aperture allocated to each function is reduced, which directly lowers beamforming gain and spatial focusing capability. This trade-off becomes particularly critical at mm-Wave frequencies, where link budgets are sensitive to aperture size and reflection efficiency.

In security-oriented scenarios, partitioned RIS architectures enable concurrent signal enhancement for legitimate users and artificial noise shaping toward adversarial directions. While experimental studies demonstrate that integrating RIS partitioning with online deep reinforcement learning–based beam selection can operate without explicit channel state information~\cite{Nasser2024Onlie}, such approaches rely on complex optimization and careful coordination across partitions. In practical deployments, limited environmental knowledge, hardware nonidealities, and dynamic channel variations increase the difficulty of real-time partition configuration and may degrade the expected secrecy gains. Beyond physical layer security, RIS partitioning has been experimentally validated as a hardware framework for supporting multiple communication functions on a single surface~\cite{Nasser2025Versatitlity}. Nevertheless, assigning different partitions to independent objectives introduces additional control overhead and coordination complexity. When the surface is partitioned into smaller sub-arrays, proof-of-concept measurement campaigns reveal fundamental trade-offs between functional diversity, effective aperture utilization, and beamforming gain. Excessive partitioning may significantly reduce spatial selectivity and diminish array-level performance.

To improve adaptability, sensing-assisted and learning-driven beam selection frameworks have been proposed for RIS-assisted systems~\cite{Kanaan2025Multimodal}. By leveraging multimodal sensing information, such as vision-based user detection and inertial measurements, deep reinforcement learning can coordinate beam selection across RIS partitions and transceivers without explicit channel estimation. However, integrating sensing modules and learning algorithms into RIS control loops introduces additional system-level challenges, including synchronization requirements, increased hardware–software coupling, and scalability constraints for large-scale deployments.
As the number of supported functions and users grows, ensuring low-latency control and stable operation across partitions remains a significant implementation challenge.

\subsection{Implementation Considerations for Emerging RIS Applications}
Emerging RIS-enabled applications extend beyond conventional communication scenarios and introduce new operational regimes that challenge existing RIS design assumptions. Large-scale RIS deployments operating in the radiative near field~\cite{Chen20246G} invalidate the widely adopted plane-wave model and require spherical-wave channel modeling and dedicated near-field signal processing~\cite{Wang2024Wideband,Zheng2023Near}. Although near-field operation enables spatial focusing and high-resolution multiplexing, it increases computational complexity and demands higher calibration accuracy, finer phase control, and tighter synchronization, making hardware nonidealities and mutual coupling effects more pronounced.

Beyond near-field regimes, RIS-assisted non-terrestrial networks (NTNs), including satellite- and UAV-based systems~\cite{Zheng2024LEO,Zhang2023Capacity}, introduce stringent constraints on power consumption, weight, reliability, and control latency, often motivating active or hybrid RIS architectures~\cite{Zhang2023Active}, which further increase hardware complexity and energy demands. The coexistence of reflective and transmissive surfaces~\cite{Zeng2021Reconfigurable}, reconfigurable antennas~\cite{zheng2025reconfigurableantennas6gtechnologies},~\cite{Ruiqi2025VTC} and integration with advanced antenna techniques such as fluid antenna systems (FAS)~\cite{Ruiqi2025ERFAS} and digital-coding metasurface arrays~\cite{liu2022programmable} enhance controllability but introduce additional coordination, interference management, and scalability challenges~\cite{Zhu2024Unified}. In addition, the antenna-in-package (AiP) paradigm, as demonstrated in~\cite{Ruiqi2024AWPL}, highlights the importance of electromagnetic–package co-design at the system integration level. Extending this packaging-oriented integration concept toward RIS-in-package implementations introduces packaging parasitics, thermal dissipation, and interconnect losses into the design loop, making electromagnetic–circuit–package co-design increasingly critical. While such highly integrated realizations improve compactness and structural robustness, they impose stricter fabrication tolerances and calibration requirements, particularly for large-scale arrays. These emerging applications tighten the coupling between electromagnetic design, control architecture, and system-level optimization, posing significant implementation challenges for large-scale and heterogeneous RIS deployments.

\section{Future Research Directions and Outlook}


While substantial progress has been made in mm-Wave RIS hardware and system modeling, the field is approaching a transition point from proof-of-concept demonstrations toward scalable and standardized deployment. In this transition, cost and power consumption are expected to become decisive constraints alongside EM performance. Future advances are therefore unlikely to be driven solely by adding new degrees of freedom, but rather by improving the efficiency, robustness, scalability, and cost–power trade-offs with which those degrees of freedom are realized.

A central direction will be the tightening of hardware–system co-design. When RIS operation expands into wideband, near-field, and sub-THz regimes, idealized reflection models become increasingly inadequate. Practical implementations should explicitly account for phase-frequency coupling, mutual coupling, calibration errors, and quantization constraints. This shift calls for hardware-aware optimization frameworks in which EM structures, biasing networks, control circuits, and signal processing algorithms are jointly designed rather than sequentially optimized. 

Moreover, cost and power consumption are expected to emerge as decisive constraints for large-scale RIS deployment. While many existing prototypes demonstrate impressive EM performance, their reliance on discrete switching components, multilayer PCB fabrication, and complex biasing networks raises concerns regarding manufacturing cost, assembly complexity, and energy efficiency. For practical 6G integration, RIS technologies should achieve significant reductions in per-element cost and static power overhead. In this context, fully-printed varactor-based RIS architectures exhibit strong potential. Fully-printed RIS can reduce fabrication cost through additive manufacturing and large-area processing, while varactor-based designs enable continuous phase tuning with improved spectral efficiency and potentially reduced switching loss. Nevertheless, these approaches introduce their own challenges, including material conductivity limitations, bias linearity, loss management, and long-term reliability. Balancing EM performance with cost-effectiveness and energy efficiency will therefore be a defining challenge for future RIS commercialization. 

In parallel, operation at higher frequencies and in heterogeneous deployment scenarios will impose stricter constraints on energy efficiency, thermal stability, fabrication precision, and integration with transceivers. Hybrid passive–active designs, slope-aware phase-frequency control, and on-chip implementations represent promising directions, yet they also amplify complexity and power-consumption challenges. Addressing these trade-offs will be critical for practical 6G integration.

Finally, the increasing convergence of RIS with integrated sensing and communication, non-terrestrial networks, fluid antenna systems, and reconfigurable antenna platforms suggests that future intelligent surfaces will not operate as isolated components, but as part of a broader reconfigurable wireless ecosystem. Developing unified control frameworks, interference-aware coordination mechanisms, and standardized performance metrics will be essential to bridge academic innovation and industrial adoption. Overall, the evolution of RIS research is expected to move from expanding theoretical capabilities toward ensuring scalable, cost-effective, energy-efficient, and system-compatible implementations. The next phase of development will likely be defined by interdisciplinary integration across electromagnetics, materials science, circuit design, signal processing, and network architecture, transforming RIS from experimental prototypes into mature and versatile infrastructure components for next-generation wireless systems.


\bibliography{references}
\bibliographystyle{IEEEtran}

\end{document}